\documentstyle[a4,psfig]{article}
\begin{document}

\title{Logarithmic corrections in the aging of the fully-frustrated Ising model}

\author{J-C. Walter and C. Chatelain\\
Laboratoire de Physique des Mat\'eriaux,\\
Universit\'e Henri Poincar\'e Nancy I,\\
BP~239, Boulevard des aiguillettes,\\
F-54506 Vand{\oe}uvre l\`es Nancy Cedex, France}
\maketitle

\begin{abstract}
We study the dynamics of the critical two-dimensional fully-frustrated Ising model
by means of Monte Carlo simulations. The dynamical exponent is estimated at
equilibrium and is shown to be compatible with the value $z_c=2$. In a second
step, the system is prepared in the paramagnetic phase and then quenched at its
critical temperature $T_c=0$. Numerical evidences for the existence of logarithmic
corrections in the aging regime are presented. These corrections may be related to
the topological defects observed in other fully-frustrated models. The
autocorrelation exponent is estimated to be $\lambda=d$ as for the Ising chain
quenched at $T_c=0$.
\end{abstract}


\def\build#1_#2^#3{\mathrel{
\mathop{\kern 0pt#1}\limits_{#2}^{#3}}}

\section*{Introduction}
Out-of-equilibrium statistical physics is a very active field of research. Among the
most studied topics, the slow evolution of glasses remains a challenging problem.
During a quench in the glass phase, the system is trapped in an infinite succession
of metastable states and never reaches equilibrium. Time-translation invariance is
broken and the fluctuation-dissipation theorem does not hold anymore~\cite{Bouchaud97,
Cugliandolo02,Crisanti02}. Frustration and randomness are two key ingredients in
the behavior of glasses. Interestingly, neither frustration nor randomness are
necessary conditions for aging. Indeed it is known for a decade that homogeneous
ferromagnets can also display aging. When they are prepared in their paramagnetic
phase and then quenched below their critical temperature $T_c$ for example,
ferromagnets age due to metastable states consisting in ferromagnetic domains in
competition. The domain walls between them cannot be eliminated in a finite
time~\cite{Bray94}. The divergence of the relaxation time provokes the suppression
of the exponential relaxation and two-time observables, for instance the autocorrelation
function $C(t,s)$, decays as a power-law of the scaling variable $t/s$. A more detailed
presentation will be given in section~\ref{sec3}.

In this work, we are interested in the intermediate situation of frustrated systems
but without disorder. The model considered is the critical two-dimensional fully-frustrated
Ising model (FFIM). The equilibrium properties of the FFIM will be presented in section
\ref{sec1}. The FFIM belongs to the same universality class as the anti-ferromagnetic
Ising model on a triangular lattice (AFIT). As a consequence, one
may assume that these two models display the same behavior out-of-equilibrium. This is
indeed claimed in a note in Ref.~\cite{Kim03}. However, controversial results are found
in the literature about the AFIT. On the one hand, numerical evidences suggesting the
existence of topological defects in anti-ferromagnetic Ising and Potts models
have been given~\cite{Kolafa84,Moore99}. Out-of-equilibrium, these topological
defects behave as impurities and pin the domain walls. The motion of the
latter is thus slowed down by a logarithmic factor which is then observable in
dynamical quantities~\cite{Bray94}. Such a mechanism is also encountered
in pure systems where topological defects are present too. The paradigmatic model in
this case is the two-dimensional XY model for which it has been shown that vortices
manifest themselves by logarithmic corrections in the long-time behavior of the
dynamical observables~\cite{Bray94,Berthier00}. Such logarithmic corrections
have been observed~\cite{Moore99} in the AFIT. Like in the XY model, the dynamical
exponent is predicted to be $z_c=2$ in the AFIT since it exists an exact
mapping onto a Gaussian Solid-on-Solid model~\cite{Nienhuis84}.  
On the other hand, a recent Monte Carlo simulation
of the anti-ferromagnetic Ising model failed to observe such logarithmic
corrections~\cite{Kim03}. In contradistinction to the previous scenario, the dynamical
exponent was estimated to be $z_c\simeq 2.33$. The spin-spin autocorrelation was
observed to decay as a power-law with an exponent $\lambda_c/z_c\simeq 0.86$.

Our goal is to investigate the existence of logarithmic corrections in the FFIM.
To that purpose, we will estimate the dynamical exponent $z_c$ at equilibrium
in section~\ref{sec2}. Remind that the two scenarii give quite different values.
In section~\ref{sec3}, we will study the aging properties and search for the
signature of topological defects when the system is quenched from the paramagnetic
phase. The ratio $\lambda/z$ will be estimated and compared to the different scenarii.

\section{The Fully-Frustrated Ising Model (FFIM)}
\label{sec1}
\begin{center}
\begin{figure}
        \centerline{\psfig{figure=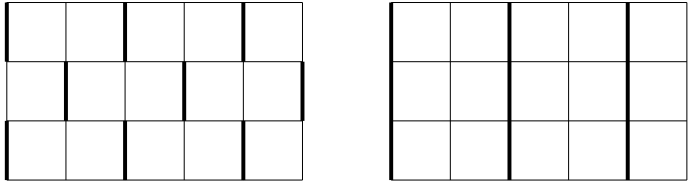,height=2.5cm}}
        \caption{Coupling configurations discussed in the text: Zig-Zag (left)
	and Piled-up Domino (right) configurations. Thin lines correspond to
	ferromagnetic couplings and thick ones to anti-ferromagnetic couplings.}
        \label{fig1}
\end{figure}
\end{center}

We consider the classical two-dimensional fully-frustrated Ising model
(FFIM) defined by the Hamiltonian
	\begin{equation}
	{\cal H}=-\sum_{x,y}\left[\sigma_{x,y}\sigma_{x+1,y}
	+(-1)^{f(x,y)}\sigma_{x,y}\sigma_{x,y+1}\right],
	\quad \sigma_{x,y}\in\{-1,1\}.
	\end{equation}
The function $f(x,y)$ is chosen such that each plaquette has an odd number
of anti-ferro\-magnetic couplings. As a consequence there exists no spin configuration
for which all bonds can be satisfied. The most studied coupling configurations
correspond to $f(x,y)=x+y$ (Zig-Zag) and $f(x,y)=x$ (Piled-up Domino). They are
depicted on Figure~\ref{fig1}. At equilibrium, these two coupling configurations
are equivalent since a transformation leaving the partition function unchanged
maps one onto the other~\cite{Mattis76}. In this study, we will consider
the Zig-Zag coupling configuration.

\begin{center}
\begin{figure}
        \centerline{\psfig{figure=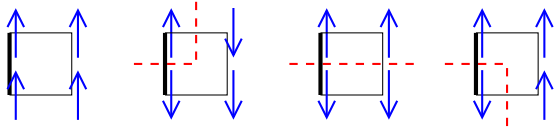,height=1.75cm}}
        \caption{Possible spin configurations on a plaquette in the
	ground state. The dashed lines correspond to the domain walls
	between the different ferromagnetic phases.}
        \label{fig2}
\end{figure}
\end{center}

\begin{center}
\begin{figure}
        \centerline{\psfig{figure=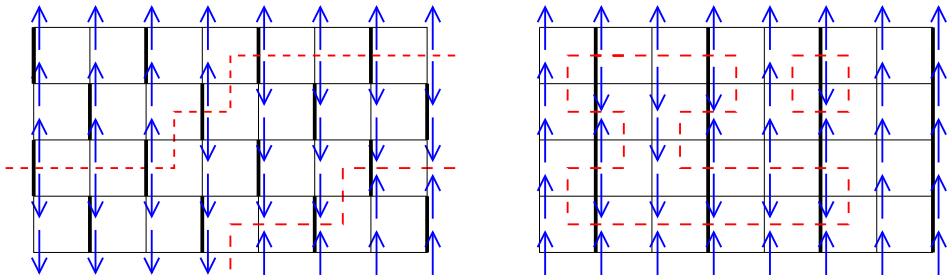,height=3.75cm}}
        \caption{Examples of spin configurations in the ground state
	of the FFIM for the Zig-Zag (left) and the Piled-up Domino coupling
	configuration (right).}
        \label{fig4}
\end{figure}
\end{center}

Spin configurations in the ground state can be obtained by minimization of the
energy of each plaquette independently. Since by construction of the model it is
not possible to satisfy all the bonds in a plaquette, the possible spin
configurations in the ground state are those with only one unsatisfied bond. They
are shown on Figure~\ref{fig2}. The ground-state is then built by juxtaposition
of these plaquettes. Even when three out of the four spins of the plaquette are
fixed by neighboring plaquettes, there may exist several possibilities to
choose the fourth spin (the two last plaquettes of Figure~\ref{fig2} give an example
of such a situation). As a consequence, the ground state is highly degenerate
and the entropy per site is finite at $T=0$~\cite{Fisher61}. One can describe the
spin configurations at $T=0$ as made of ferromagnetic domains separated by
domain walls. The possible plaquettes in the ground state forbid that more than
one domain wall goes through a plaquette which means that domain walls cannot
intersect each other. Moreover, when going through a plaquette, the domain wall
must cross the anti-ferromagnetic bond and one of the three ferromagnetic bonds.
Figure~\ref{fig4} shows examples of spin configurations in the ground state.
In the case of the Zig-Zag coupling configuration, domain walls cannot form
overhangs or loops. They must start and end at two different boundaries of the system.
For the Piled-up Domino coupling configuration, domain walls can form overhangs
and loops. They can thus enclose finite clusters that can be as small
as one single spin. 

The equilibrium properties of the FFIM have been determined by analytical
diagonalization of the transfer matrix~\cite{Villain77}. The ferromagnetic
order is destroyed at any temperature because the domain walls discussed
above have no energy. Nevertheless, the system is critical at $T_c=0$
with spin-spin correlation functions decaying algebraically with an exponent
$\eta=1/2$~\cite{Forgacs80}. As already mentioned in the introduction, the
FFIM belongs to the same universality class as the homogeneous anti-ferromagnetic
Ising model on a triangular lattice~\cite{Wannier50} (AFIT). 

\section{Dynamical exponent of the FFIM}
\label{sec2}
We now consider the dynamical properties of the FFIM at equilibrium at its
critical temperature $T_c=0$. The system is initially thermalized and then
evolves with the Glauber dynamics~\cite{Glauber63}. For an inverse temperature
$\beta=1/k_BT$, the probability for a spin to be flipped is ${e^{-\beta\Delta E}
\over 1+e^{-\beta\Delta E}}$ where $\Delta E$ is the variation of energy
when the spin is flipped. In our case, the temperature is $T=0$ which means that
a spin flip is always accepted if the energy is lowered, accepted with a
probability $1/2$ if the energy does not change, and always rejected otherwise.
Moreover, we start from an equilibrium state at $T=0$, i.e. with the minimal
value of the energy. As a consequence, the algorithm becomes particularly simple:
a spin is randomly chosen and is flipped if the total energy does not
change~\footnote{Changing the probability of a spin flip from $1/2$ to $1$
leads here only to a factor of $2$ in the autocorrelation time.}.

In contradistinction to homogeneous ferromagnets, the Glauber dynamics of
the FFIM is not frozen at $T=0$. In the case of the Zig-Zag coupling configuration,
some single-spin flips, as for instance the formation of a dot on a flat domain
wall, does not change the total energy and are thus allowed. In the case of the
Pile-up Domino coupling configuration, one can also flip any spin which lies on
an anti-ferromagnetic bond in the bulk of a ferromagnetic domain. This process
creates a new single-spin domain as the one depicted on Figure~\ref{fig4}.
These single-spin domains can be flipped without any energy change. As a
consequence, they behave as isolated Ising spins. This situation was already
encountered in the case of the AFIT where these spins were called
{\sl loose spins}~\cite{Forgacs80}.

The dynamical exponent $z_c$ of the FFIM at $T=0$ is not known. Since the FFIM belongs
to the same universality class at equilibrium as the AFIT, one may assume that they
share the same dynamical exponents $z_c$. As already mentioned in the introduction,
different values can be found in the literature. According to Moore
{\sl et al.}~\cite{Moore99} who claim the existence of topological defects,
the dynamical exponent should be $z_c=2$. On the other hand, the numerical estimate
$z_c\simeq 2.33$ has been determined in more recent Monte Carlo simulations~\cite{Kim03}. 

\begin{center}
\begin{figure}
        \centerline{\psfig{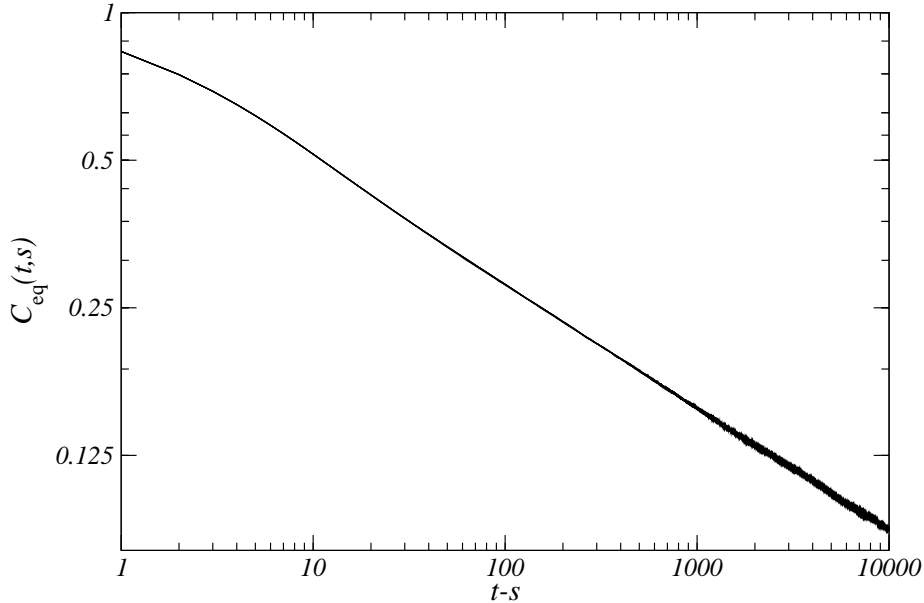}}
        \caption{Equilibrium spin-spin autocorrelation function $C(t,s)$ of the FFIM
	at $T_c=0$ with the Zig-Zag coupling configuration versus $t-s$. The six curves
	corresponding to $s=10,20,40,160,640$ and $s=1280$ cannot be distinguished
	as expected at equilibrium. A very nice power-law is observed and allows
	for a precise determination of $2\beta/\nu z_c$.}
        \label{fig5}
\end{figure}
\end{center}

To measure the critical exponent in the FFIM, we computed the spin-spin
autocorrelation function at equilibrium at $T_c=0$. The lattice size is $192\times
192$ and we used periodic boundary conditions in both directions. The system is
first thermalized using the Kandel-Ben~Av-Domany cluster algorithm~\cite{Kandel92}.
We then use the Glauber dynamics and measure $\sigma_i(s)$ and $\sigma_i(t)$.
The simulation is repeated 1000 times. The mean spin-spin autocorrelation is
finally estimated as
	\begin{equation}
	C_{\rm eq.}(t,s)={1\over N}\sum_i\langle \sigma_i(t)\sigma_i(s)\rangle
	\end{equation}
where $\langle\ldots\rangle$ denotes the average over the 1000 histories of the
system. As shown on Figure~\ref{fig5}, the correlation function decays as a power-law
for sufficiently large separation time ($t-s>10$). One indeed expects that the
divergence of the relaxation time at the critical point leads to an algebraic
decay of the autocorrelation function $C(t,s)$. Moreover, time-translation invariance
at equilibrium imposes a dependence on $t-s$. Finally, the order parameter having an
anomalous dimension $\beta/\nu$, one expects
	\begin{equation}
	C_{\rm eq.}(t,s)\build\sim_{t-s\gg 1}^{} (t-s)^{-2\beta/\nu z_c}
	\end{equation}
The interpolation of the data (Figure~\ref{fig5}) with this expression leads to
the estimate $2\beta/\nu z_c=0.2495(11)$. Since $\eta=2\beta/\nu=1/2$ ($d=2$), our
estimate for the dynamical exponent is $z_c\simeq 2.004(9)$. This value is
compatible with $z_c=2$, i.e. the one expected in the scenario involving topological
defects~\cite{Moore99} and contradicts the observation of a sub-diffusive
growth with $z\simeq 2.33$ observed by Kim {\sl et al.}~\cite{Kim03} with
random initial configurations in the AFIT.
Note that we also measured the dynamical exponent $z_c$ during a quench 
by considering the decay of autocorrelation functions in the quasi-equilibrium
regime. We obtained again a value compatible with $z_c=2$.

\section{Aging of the FFIM}
\label{sec3}
\begin{center}
\begin{figure}
        \centerline{\psfig{figure=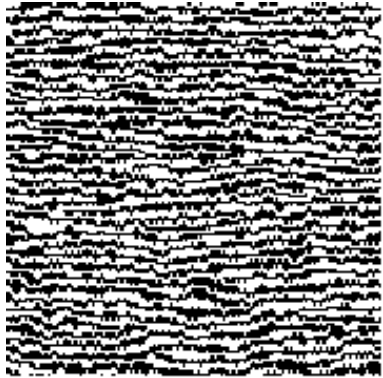,height=5cm}}
        \caption{Typical spin configuration of the FFIM with the Zig-Zag coupling
	configuration when initially prepared at $T\rightarrow +\infty$ and then
	quenched at $T_c=0$.}
        \label{fig3}
\end{figure}
\end{center}

We are now interested in the out-of-equilibrium dynamics of the FFIM. The system is initially
prepared at $T\rightarrow +\infty$, i.e. spins are random. It it then quenched at the critical
temperature $T_c=0$. We use the Glauber dynamics at $T_c=0$: a spin flip is always
accepted when it leads to a decrease of the total energy ($\Delta E<0$) and only with a
probability $1/2$ if $\Delta E=0$. A typical spin configuration is presented on
Figure~\ref{fig3}. As expected, ferromagnetic domains grow during the quench and
the domain walls tend to the stable configurations discussed in section~\ref{sec1}.
On the snapshot (Figure~\ref{fig3}), most of the domains are already in the ground
state. A few others do not span the entire system and are thus unstable. One expects
that they will disappear by shrinking or moving to one of the boundaries.
In the following, we will show that the aging of the FFIM can be analyzed using the same
assumptions as for homogeneous ferromagnets.

In the case of homogeneous ferromagnets quenched below their critical temperature,
domains grow with a characteristic length $L(t)$ behaving in time as $L(t)\sim t^{1/z}$
where $z=2$ is the dynamical exponent~\cite{Hohenberg77}. The two-time autocorrelation
function of the local order parameter can be decomposed as
	\begin{equation}
	C(t,s)=C_{\rm st.}(t-s)+C_{\rm ag.}(t/s)
	\end{equation}
where $C_{\rm st.}(t-s)$ is due to reversible processes occurring inside the
domains while $C_{\rm ag.}$ comes from the irreversible motion and annihilation
of domain walls. This decomposition is clearly seen for example in the case of
the spherical model~\cite{Godreche00b}. Usually, the first term falls down rapidly
and can be neglected. The self-similarity of the domain growth implies that
two-time observables, say $C(t,s)$, depend only on $L(t)/L(s)$ and thus on $t/s$
when $1\ll s\sim t$, i.e. in the aging regime~\cite{Bray94}. The autocorrelation
function $C(t,s)$ behaves as~\cite{Janssen89,Godreche01}:
	\begin{equation}
                C_{\rm ag.}(t,s)\sim M_{\rm eq}^2f_C\left({t\over s}\right)
        	\label{ScalingCR}
        \end{equation}
with the scaling variable $t/s$. In the limit $t\gg s$, the scaling function
$f_C(x)$ decays algebraically as $x^{-\lambda/z}$. This defines a new exponent,
the autocorrelation exponent $\lambda$~\cite{Fisher88}. At the critical point,
similar conclusions can be drawn from the assumption that the correlation length
grows during a quench as $\xi\sim t^{1/z_c}$. The autocorrelation function
$C(t,s)$ displays a scaling behavior analogous to (\ref{ScalingCR}):
	\begin{equation}
                C_{\rm ag.}(t,s)\sim s^{-a_c}g_C\left({t\over s}\right)
        	\label{ScalingCR2}
        \end{equation}
where $g_C(x)$ is expected to decay asymptotically as $x^{-\lambda_c/z_c}$.
Dynamical and autocorrelation exponents $z_c$ and $\lambda_c$ take values different
from below $T_c$. Because $C(s,s)=\langle M^2(s)\rangle\sim \xi^{-2\beta/\nu}\sim
s^{-2\beta/\nu z_c}$, the exponent $a_c$ is equal to ${2\beta/\nu z_c}$.
The expression (\ref{ScalingCR2}) works for a broad range of models for example
the spherical model~\cite{Godreche00b} or the Ising chain~\cite{Godreche00a}
that can be both treated analytically, the 2D Ising model by Monte Carlo
simulations~\cite{Godreche00b} or the $O(n)$ model analyzed by Renormalization Group
technics~\cite{Calabrese04}. In the case of the XY model, the motion of the domains
is slowed down by the vortices that behave as impurities pinning the domain walls.
The correlation length grows as $(t/\ln t)^{1/z_c}$ so that the autocorrelation
function is modified by the replacement of the scaling variable $t/s$ by
$t\ln s/s\ln t$~\cite{Bray94b,Bray94}:
	\begin{equation}
                C_{\rm ag.}(t,s)\sim s^{-a_c}\tilde g_C\left({t\ln s\over s\ln t}\right)
        	\label{ScalingCR3}
        \end{equation}

We will analyze the aging of the FFIM in two steps. First, we will determine whether
topological defects exist in the FFIM and affect the dynamics as claimed by some authors
in the case of the AFIT. To that purpose, we will determine which one of the two
scenarii (\ref{ScalingCR2}) or (\ref{ScalingCR3}) is the more compatible with our data,
i.e. which scaling variable, $t/s$ or $t\ln s/s\ln t$, leads to a collapse of the scaling
function $s^{a_c}C(t,s)$ for different waiting times $s$. Then, we will estimate the
autocorrelation exponent $\lambda$ using the appropriate scaling variable. For these
simulations, the lattice size is again $L=192$ with periodic boundary conditions but
the data have been averaged over 50,000 different histories of the system.

\subsection{Determination of the scaling variable}
\begin{center}
\begin{figure}
        \centerline{\psfig{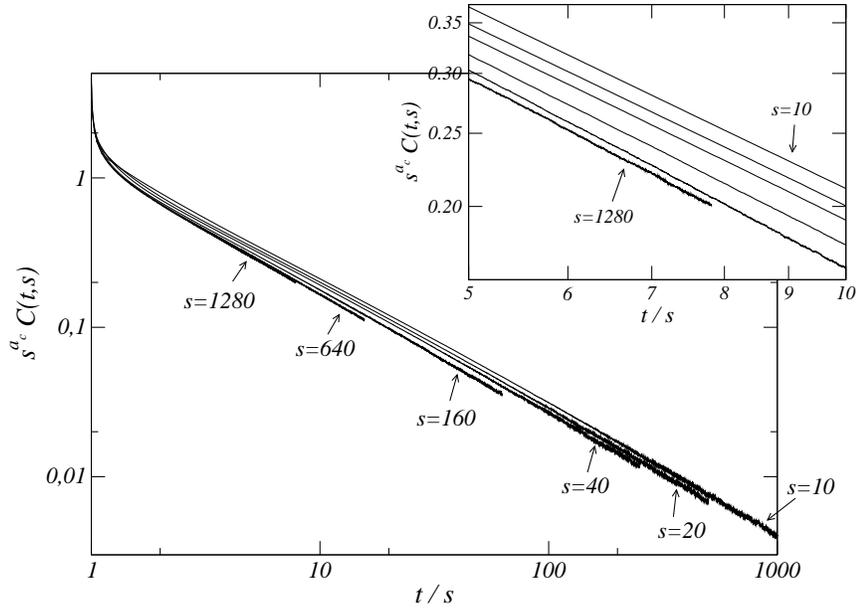}}
        \caption{Scaling function $s^{a_c}C(t,s)$ of the spin-spin autocorrelation
	function versus $t/s$. The trend does not show any real trend to scaling.}
        \label{figScal}
\end{figure}
\end{center}

\begin{center}
\begin{figure}
        \centerline{\psfig{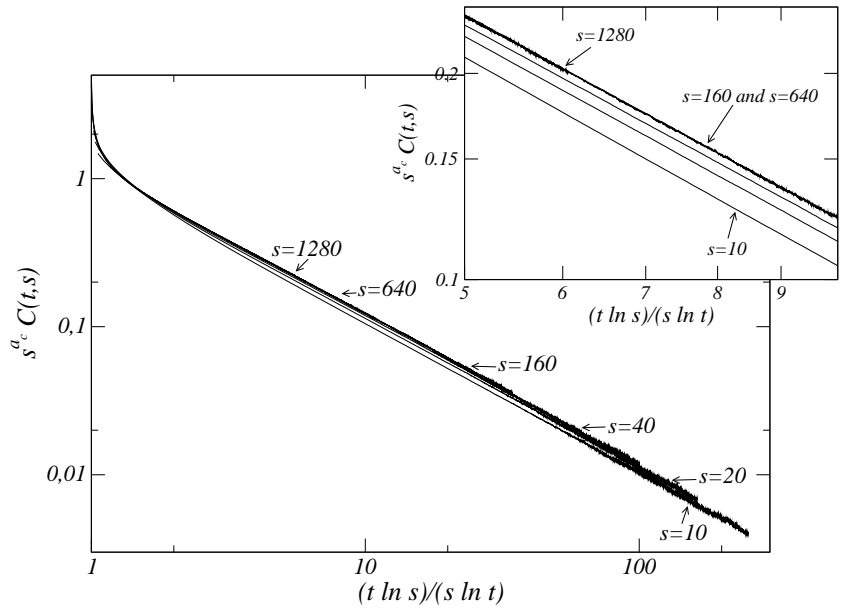}}
        \caption{Scaling function $s^{a_c}C(t,s)$ of the spin-spin autocorrelation
	function versus $t\ln s/s\ln t$. A perfect collapse of the curve for the
	three largest waiting times can be observed in the inset.}
        \label{figScal-log}
\end{figure}
\end{center}

Our numerical estimation of the scaling function $s^{a_c}C(t,s)$ is
plotted with respect to $t/s$ on Figure~\ref{figScal} and to $t\ln s
/s\ln t$ on Figure \ref{figScal-log}. We used the exponent
$a_c=0.2495(11)$ obtained in section~\ref{sec2}. No collapse of the
scaling function $s^{a_c}C(t,s)$ is observed with the scaling variable
$t/s$ (Figure~\ref{figScal}). The last two curves ($s=640$ and $1280$)
are however closer to each other. Our waiting times $s$ are too small to
check if this is the sign of an accumulation of the curves for much larger
waiting times. In contradistinction, a very good collapse for large
waiting times $s\ge 160$ can be seen in the inset of Figure~\ref{figScal-log}
when the data are plotted versus $t\ln s/s\ln t$. The difference between
the different curves is of the order of $2.10^{-3}$
where the correlation itself is of order of $0.2$. The difference between
the different curves is of the order of magnitude of the statistical
fluctuations of the data. For these reasons, it is unlikely that this
collapse for the three waiting times $s=160$, $s=640$ and $1280$ be
accidental and disappear for larger waiting times. We have also tried more
complex scaling variables as $x=t(\ln s)^\mu/s(\ln t)^\mu$ or $x=t\ln {s\over t_0}
/s\ln {t\over t_0}$ but they do not lead to a better collapse.

Note that the observation of these logarithmic corrections has been made
possible by the use of the two times $t$ and $s$. It is well known that it is
very difficult to distinguish these logarithmic corrections if ones studies
only the asymptotic dependence of $C(t,s)$ on $t$, i.e. $C(t,s)\sim
(t/\ln t)^{-\lambda/z}$. Moreover, if such corrections are observed, they
may also result from corrections to scaling, i.e. $C(t,s)\sim t^{-\lambda/z}
(1+at^{-\omega}+\ldots)$ with $\omega$ small. The relations (\ref{ScalingCR2})
or (\ref{ScalingCR3}) imposes a much stronger constraint than the asymptotic
law $C(t,s)\sim (t/\ln t)^{-\lambda/z}$. We are looking for logarithmic
factors in the scaling variable. The analysis is thus not sensitive to
scaling corrections to the dominant algebraic decay. These considerations
may explain why these logarithmic corrections, although observed by~\cite{Moore99}
for the AFIT have not be seen in more recent Monte Carlo simulations~\cite{Kim03}.

\subsection{Determination of the exponent $a_c$}
The previous analysis relies strongly on the scaling form (\ref{ScalingCR}).
Our conclusion may be criticized by arguing that corrections to the prefactor
$s^{-a_c}$ should have been taken into account. To show that these corrections
are negligible for the range of times $s$ and $t$ where we observe the collapse
of the scaling function, we studied the two-time correlation $C(t,s)$ on the family
of curves $t\ln s/t\ln s=\kappa$ where $\kappa$ is a constant. Along these curves,
the scaling function $g_C(\kappa)$ is constant and the scaling form
(\ref{ScalingCR}) reduces to $s^{-a_C}$. We then defined an effective
exponent as
	\begin{equation}
	a_{\rm eff}(s_1,s_2)=-{\ln C(t_2,s_2)-\ln C(t_1,s_1)
	\over\ln s_2-\ln s_1}
	\end{equation}
where $t_1\ln s_1/t_1\ln s_1=t_2\ln s_2/t_2\ln s_2=\kappa$. In practice,
to allow for non-integer values of $t_1$ and $t_2$ the logarithm of the
autocorrelation function $C(t,s)$ was interpolated linearly between the
two nearest simulation points, i.e. $\ln C(t,s)=(t-{\rm Int}\ \!t)
\ln C({\rm Int}\ \!t,s)+(1-t+{\rm Int}\ \!t)\ln C({\rm Int}\ \!t+1,s)$ where
${\rm Int}\ \!t$ is the largest integer smaller or equal to $t$.

\begin{center}
\begin{figure}
        \centerline{\psfig{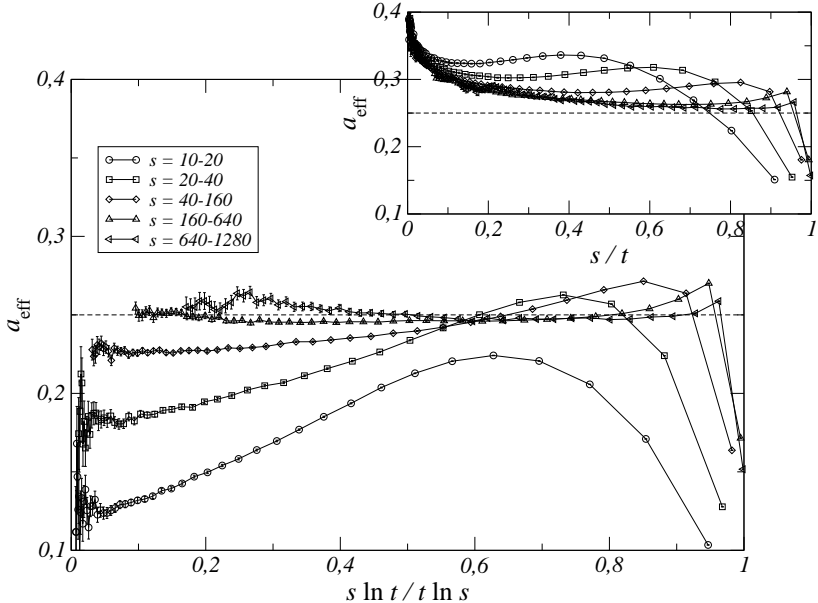}}
        \caption{Effective exponent $a_{\rm eff}$ estimated by a fit of the
	correlation function as $C(t,s)\sim s^{-a_c}$. The time $t$ is chosen such 
	that the scaling variable $t\ln s/s\ln t$ (or $t/s$ in the inset),
	and thus the scaling function $\tilde g_C$ in Eq.~\ref{ScalingCR3}
	($g_C$ in Eq.~\ref{ScalingCR2}), is kept constant. The legend indicates
	the two waiting times $(s_1,s_2)$ used for the calculation of the
	effective exponent.}
	\label{fig6}
\end{figure}
\end{center}

The effective exponent is expected to tend to the exact value $a_c=0.25$
in the limit of large waiting times $s$. At finite values of $s$, corrections
to the behavior $s^{-a_c}$ should lead to effective exponents that differ
from $a_c=0.25$. Our data are plotted on Figure \ref{fig6}. One can see
that the effective exponent displays as expected a plateau around the
exact value $a_c=0.25$ when sufficiently large waiting times
$s\ge 160$ are used. The same analysis have been performed along the curve $t/s=\kappa
={\rm Cst}$. It leads to an effective exponent significantly larger than the exact
value $a_c=0.25$. Moreover the absence of any plateau of the effective exponent
excludes the possibility of a value of $a_c$ different from $0.25$.
This analysis confirms that the scaling form that is the most compatible
with our data is (\ref{ScalingCR3}) and not (\ref{ScalingCR2}).

\subsection{Determination of the autocorrelation exponent $\lambda_c$}
\begin{center}
\begin{figure}
        \centerline{\psfig{figure=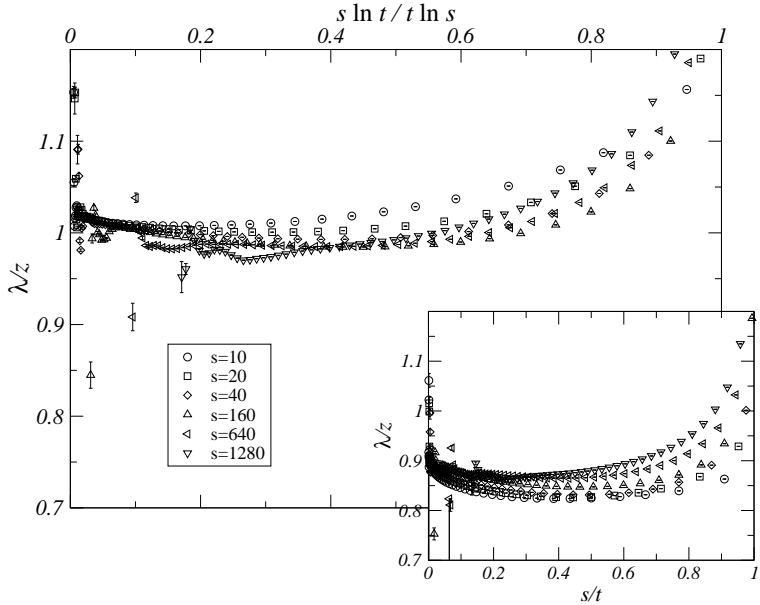,height=8cm}}
        \caption{Effective exponent $\lambda_c/z_c$ obtained by a
	power-law fit $C(t,s)\sim x^{-\lambda_c/z_c}$ where $x$ is the scaling
	variable, i.e. $x=t\ln s/s\ln t$ (main figure) when taking into
	account topological defects and $x=t/s$ (insert) otherwise. The fit
	is performed in the range $x\in [x_{\rm min};x_{\rm max}]$ where
	$x_{\rm max}$ is the maximum value of the scaling variable allowed by
	our data. $x_{\rm min}$ is varied and the effective exponent $\lambda_c/z_c$
	is plotted with respect to $1/x_{\rm min}$. Since we are interested
	in the asymptotic regime where corrections to scaling can be neglected,
	the interesting region corresponds to $x_{\rm min}$ large.}
        \label{fig8}
\end{figure}
\end{center}

We are now interested in the autocorrelation exponent $\lambda_c$ defined
by the assumption that the scaling function $\tilde g_C$ decays asymptotically
as $\tilde g_C(x)\sim x^{-\lambda_c/z_c}$ where $x=t\ln s/s\ln t$ is the scaling
variable. To take into account the possibility of corrections to this
behavior, we studied the effective exponent $(\lambda_c/z_c)_{\rm eff.}
(x_{\rm min},x_{\rm max})$ defined as the decay exponent of the
autocorrelation function $C(t,s)$ with $x=t\ln s/s\ln t$ in the window
$x\in[x_{\rm min};x_{\rm max}]$. The upper bound $x_{\rm max}$ is the largest
scaling variable allowed by our data, i.e. $x_{\rm max}=x(s,t=10000)$. The
low bound of the interpolation range is varied between $x(s,s)=1$ and
$x_{\rm max}$. The data are plotted with respect to $1/x_{\rm min}$ on
Figure~\ref{fig8}. For small values of $x_{\rm min}$, almost all numerical
data, including small values of $x$, are included in the power-law interpolation.
As a consequence, the corrections may play an important r\^ole and lead to
an effective exponent possibly quite different from the asymptotic value $\lambda_c/z_c$.
When $x_{\rm min}$ is increased, the effective exponent is expected to tend
to $\lambda_c/z_c$. When $x_{\rm min}$ approaches $x_{\rm max}$, the number
of points that contribute to the fit decreases and the effective exponent
becomes noisier. The exponent $\lambda_c/z_c$ has to be estimated in an
intermediate regime.

The data are plotted on Figure~\ref{fig8}. For large values of $x_{\rm min}$, the
effective exponent stays inside a range corresponding to $\lambda_c/z_c\simeq
1.02(2)$. This means that the autocorrelation exponent is $\lambda_c\simeq 2.04(5)$.
and therefore appears to be saturated, i.e. $\lambda_c=d$. The ratio $\lambda_c/z_c$
remains compatible with our estimate $1.02(2)$ when considering more complex scaling
variables involving logarithmic corrections as $x=t(\ln s)^\mu/s(\ln t)^\mu$ or
$x=t\ln {s\over t_0}/s\ln {t\over t_0}$. The same procedure with the scaling variable
$x=t/s$ leads to an incompatible estimate $(\lambda_c/z_c)_{\rm eff}\simeq 0.91(2)$.
This difference may be due to the fact that the asymptotic regime is seen only
for larger times if one does not take into account logarithmic corrections.
Note that the decay exponent $0.86$ was obtained for the AFIT when no logarithmic
correction is considered~\cite{Kim03}. One can thus assume that taking into account
these corrections, an exponent closer to $1$ would have been obtained.

\section{Conclusions}
We have given numerical evidences in favor of the existence of topological defects
in the paramagnetic phase of the FFIM. The dynamical exponent at the critical
temperature $T_c=0$ is estimated to be $z=2.004(9)$, in excellent agreement with the value
$z=2$ expected for the AFIT in the scenario involving topological defects~\cite{Moore99}
and fully incompatible with the Monte Carlo study where the signature of these
topological defects was not seen. We have then studied the decay of the spin-spin
autocorrelation function $C(t,s)$. Assuming that $C(t,s)$ scales as the homogeneous
XY ferromagnet, a good collapse of the scaling function is observed for large
waiting times. Moreover, the exponent $a_c$ is correctly obtained. This would not be
the case if logarithmic corrections were not taken into account.
Interestingly, the autocorrelation exponent is compatible with the simple
value $\lambda_c=d$. Note that it is also the case for the Ising chain whose
critical temperature is $T_c=0$ like the FFIM. 

An important question remains unsolved: in the AFIT, the topological defects are
known to be vortices. The couplings are not homogeneous in the FFIM so it is not
obvious to us how these topological defects look like for this model. Unfortunately,
they cannot be observed on snapshots of the spin configurations. Note that it is
also the case for the AFIT. Moore {\sl et al.} managed to pin vortices and
make it observable by averaging over a large number of spin configurations. Without
an idea of the shape of these topological defects in the FFIM, we have been unable to
apply this procedure to the FFIM. Further investigation in this direction is necessary.

\section*{Acknowledgments}
The laboratoire de Physique des Mat\'eriaux is Unit\'e Mixte de Recherche
CNRS number 7556. The authors gratefully thank the Statistical Physics group in Nancy
and especially Bertrand Berche for a carefull reading of the manuscript and Ferenc
Igl\'oi for interesting discussions and advices.


\end{document}